\documentstyle[times]{article}

\begin{document} 

\begin{titlepage}

\title{Integral Octonions, Octonion XY-Product, and the Leech Lattice}

\author{Geoffrey Dixon\\ 
 www.7stones.com \\
}

\maketitle

\begin{abstract}
The integral octonions arise from the octonion XY-product.  A parallel is shown to exist 
with the quaternion Z-product.  Connections to the laminated lattices, 
$\Lambda_{4}$, $\Lambda_{8}$, $\Lambda_{16}$ and $\Lambda_{24}$ (Leech), are developed.
\end{abstract}

\end{titlepage}

\section*{1. Octonion Multiplication and XY-Product Variants.}

Given any multiplication table for the 8-dimensional octonion algebra, \textbf{O},
one can construct an infinite number of variants, isomorphic to \textbf{O} itself, by  
replacing the original product with the XY-product,
%
%
\begin{equation}
A\circ_{XY}B \equiv (AX)(Y^{\dagger}B), 
\end{equation}
 where $YX^{\dagger}$ must be a unit octonion, and is the identity of the new 
 multiplication.  The octonion identity is left 
unchanged in the case $Y^{\dagger} = X^{-1}$, giving rise to the X-product:
%
%
\begin{equation}
A\circ_{X}B \equiv (AX)(X^{-1}B).
\end{equation}

in what follows, let $e_{a}$, a = 0,...,7, represent the 8 octonion units, with 
$e_{0} = 1$ the identity (in \textbf{[1,2]}  
the index $\infty$ is used for the identity, but the index choice 0  
makes it easier to program computers to do calculations). Our starting 
multiplication table is the commonly chosen cyclic multiplication: 
%
%
\begin{equation}
e_{a}e_{a+1}=e_{a+3}, \; \; a = 1,...,7,
\end{equation}
where the indices are taken modulo 7, from 1 to 7.   
This particular table is invariant with respect to both index cycling and index doubling.  
That is,
$$
\begin{array}{l}
e_{a}e_{b} = e_{c} \Longrightarrow e_{a+1}e_{b+1} = e_{c+1}, \\
e_{a}e_{b} = e_{c} \Longrightarrow e_{2a}e_{2b} = e_{2c}. \\
\end{array}
$$

\section*{2. $\Lambda_{8}$ Lattices.}

Representations of the $E_{8} = \Lambda_{8}$ lattice arise from 
the X-product.  In particular, define
%
%
\begin{equation}
\begin{array}{ll}
\Xi_{0} = &\{\pm e_{a}\},  \\ 
\Xi_{2} = &\{(\pm e_{a}\pm e_{b}\pm e_{c}\pm e_{d})/2: a,b,c,d
\mbox{ distinct},   \\ 
&e_{a}(e_{b}(e_{c}e_{d}))=\pm 1\}, \\ 
\Xi^{even} =&\Xi_{0}\cup \Xi_{2}, \\ 
{\cal E}_{8}^{even} =&\mbox{span\{}\Xi^{even}\mbox{\}}, \\ \\

\Xi_{1} = &\{(\pm e_{a}\pm e_{b})/\sqrt{2}: a,b \mbox{
distinct}\},  \\ \\
\Xi_{3} = &\{(\sum_{a=0}^{7}\pm e_{a})/\sqrt{8}:
\mbox{ even number of +'s}  \},   \\ 
&a,b,c,d\in\{0,...,7\},  \\ 
\Xi^{odd} =&\Xi_{1}\cup \Xi_{3}, \\ 
{\cal E}_{8}^{odd} =&\mbox{span\{}\Xi^{odd}\mbox{\}} \\ 
\end{array}
\end{equation} \\
(spans over the integers, {\cal Z}).  
$\Xi^{even}$ has 16 + 224 = 240 elements, and $\Xi^{odd}$ has 112 + 128 = 240 
elements, each a representation of the inner shell of an $E_{8}$ lattice.  (One may think 
of these $\Lambda_{8}$ lattices as discrete versions of $S^{7}$, the algebra of unit octonions.)

These elements have an interesting relation to our chosen octonion multiplication: 
for all $X \in \Xi^{even} \cup \Xi^{odd}$, and for all pairs of 
octonion units $e_{a}$, $e_{b}$, there exists a unit $e_{c}$ such that 
%
%
\begin{equation}
e_{a}\circ_{X}e_{b} = \pm e_{c}.
\end{equation}

\section*{3. Integral Octonions and the X-Product.}

Define 
%
%
\begin{equation}
\ell_{0} = \frac{1}{2}(1+e_{1}+e_{2}+e_{3}+e_{4}+e_{5}+e_{6}+e_{7}).
\end{equation} \\
Note that $\ell_{0}$, like our multiplication table, is invariant with respect 
to index cycling and doubling.  
Therefore, the X-product
%
%
\begin{equation}
A\circ_{\ell_{0}}B = (A\ell_{0})(\ell_{0}^{-1}B)
\end{equation}
is also invariant with respect to index cycling and doubling.  Its multiplication 
table is in some sense dual to that given above:
$$
e_{a}\circ_{\ell_{0}}e_{a+2}=e_{a+3}, \; \; a = 1,...,7.
$$

What about the X-product
%
%
\begin{equation}
A\circ_{\ell_{0}^{-1}}B = (A\ell_{0}^{-1})(\ell_{0}B)?
\end{equation}
Since $\ell_{0}^{-1} = \frac{1}{4}(1-e_{1}-...-e_{7})$ has an odd number of plus signs 
(as coefficients of the units), we don't expect the product of any two units to be 
another unit using this X-product.  For example,
$$
e_{1}\circ_{\ell_{0}^{-1}}e_{2} = \frac{1}{2}(e_{3} - e_{5} + e_{6} + e_{7}).
$$
While the value on the righthand side of this equality is not an octonion unit, it is 
an element of ${\cal E}_{8}^{even}$, as are $e_{1}$ and $e_{2}$.  This is a specific example 
of a more general result given below.

In section 5 we look at the laminated lattice $\Lambda_{4}$ over the quaternion algebra \textbf{H}.  
This algebra is known to be closed with respect to quaternion multiplication, giving rise to the 
algebra of integral quaternions.  However, surprisingly, the $\Lambda_{8}$ lattice ${\cal E}_{8}^{even}$ 
does not close with respect to our given octonion multiplication.  Weirdly, the set $\Xi^{even}[0-a]$, derived 
from $\Xi^{even}$ by replacing each occurrence of $e_{0}$ in elements of $\Xi^{even}$  with $e_{a}$, and 
vice versa, is multiplicatively closed.  These representations of unit integral octonions are also derivable 
from the promised general result. \\ \\
\textbf{X-Product Integral Octonion Result}
%
%
\begin{equation}
\begin{array}{l}
X \in \Xi^{odd}, \mbox{ and } A,B \in \Xi^{even} \Longrightarrow 
 A\circ_{X^{\dagger}}B \in \Xi^{even}.
\end{array}
\end{equation}
\textbf{Proof:} \\
To prove (9) we will set up a partial multiplication table for the $\Xi_{m}$ (since these sets are 
finite, proving these results even with a computer is not difficult).  In general,
$$
A \in \Xi^{even}, \mbox{ and } Y \in \Xi^{odd} \Longrightarrow YA \in \Xi^{odd}, AY^{\dagger} \in \Xi^{odd\dagger},
$$
and 
$$
X, Y \in \Xi^{odd} \Longrightarrow Y^{\dagger}X \in \Xi^{even}.
$$
(Interestingly these results are not commutative.  For example, 
$X \in \Xi_{1}, \mbox{ and } Y \in \Xi_{3} \Longrightarrow YX, XY^{\dagger} \in \Xi^{even}[0-a], (a \ne 0)$.)  
Therefore, 
$$
X \in \Xi^{odd}, \mbox{ and } A,B \in \Xi^{even} \Longrightarrow 
AX^{\dagger} \in \Xi^{odd\dagger} \mbox{ and } XB \in \Xi^{odd} \Longrightarrow 
A\circ_{X^{\dagger}}B = (AX^{\dagger})(XB) \in \Xi^{even}.
$$
That is, $\Xi^{even}$, which is not multiplicatively closed with respect to the given octonion product, 
is closed with respect to these X-products. $\Box$ \\

As a corollary, given the identity 
$$
A(BX) = (A\circ_{X^{\dagger}}B)X.
$$
we see that $\Xi^{even}$ closes as a set of actions on each $X \in \Xi^{odd}$, and more generally on $\Xi^{odd}$ itself 
\textbf{[10]}.  Note that of all the $\Xi_{k}$, only $\Xi_{3}$ is not invariant with respect to octonion conjugation.

\section*{4. Integral Octonions, the XY-Product and $\Lambda_{16}$ Lattices.}

The set of $X, Y \in \textbf{O}$ satisfying the property that for all units $e_{a}$ and $e_{b}$ there exists some unit 
$e_{c}$ such that,
%
%
\begin{equation}
e_{a}\circ_{XY}e_{b} = (e_{a}X)(Y^{-1}e_{b}) = \pm e_{c},
\end{equation}
gives rise to two copies of $\Lambda_{16}$ \textbf{[6]}.  In particular, (10) is satisfied if 
$X \in \Xi^{even} \cup \Xi^{odd}$ and there exists some unit $e_{d}$ such that $Y = \pm e_{d}X$.  
In the case $Y^{-1} = X^{-1}e_{d}^{\dagger}$ one can show using X-product identities that for 
all $A, B \in \textbf{O}$,
%
%
\begin{equation}
(AX)((X^{-1}e_{d}^{\dagger})B) = A \circ_{X} (e_{d}^{\dagger} \circ_{X} B).
\end{equation}
If $X^{-1} \in \Xi^{odd}$, then $\Xi^{even}$ is closed with respect to X-product 
multiplication.  If $A, B \in \Xi^{even}$, then $C = (e_{d}^{\dagger}\circ_{X}B) \in \Xi^{even}$, 
so $A\circ_{X}C \in \Xi^{even}$.  That is,
$$
A, B \in \Xi^{even} \mbox{ and } X^{-1} \in \Xi^{odd} \Longrightarrow (AX)((X^{-1}e_{d}^{\dagger})B) \in \Xi^{even}.
$$
So ${\cal E}_{8}^{even}$ are octavian integers with respect to this XY-product.  The identity 
of this set of integers is just $e_{d}$.

But (9) can clearly be generalized even further: \\ \\
\textbf{XY-Product Integral Octonion Result}
%
%
\begin{equation}
A,B \in \Xi^{even} \mbox{ and } X,Y \in \Xi^{odd\dagger} \mbox{ and } |XY^{-1}| = 1 
\Longrightarrow A\circ_{XY}B \in \Xi^{even}.
\end{equation}
\textbf{Proof:} \\
In general, and on the assumption my computer code was without error:
$$
\begin{array}{l}
\Xi^{even}\Xi^{odd\dagger} = \Xi^{odd\dagger}, \\
\Xi^{odd}\Xi^{even} = \Xi^{odd}, \\
\Xi^{odd\dagger}\Xi^{odd} = \Xi^{even}. \\
\end{array}
$$
Therefore, generalizing the results above:
$$
\begin{array}{l}
(\Xi^{even}\Xi^{odd\dagger}) (\Xi^{odd}\Xi^{even}) = \Xi^{odd\dagger}\Xi^{odd} = \Xi^{even}, \\
(\Xi^{odd}\Xi^{even})(\Xi^{odd\dagger}\Xi^{odd}) = \Xi^{odd}\Xi^{even} = \Xi^{odd}, \\
(\Xi^{odd\dagger}\Xi^{odd})(\Xi^{even}\Xi^{odd\dagger}) = \Xi^{even}\Xi^{odd\dagger} = \Xi^{odd\dagger}. \\
\end{array}
$$
So there exist XY-products under which each of these lattice inner shells, $\Xi^{even}$, $\Xi^{odd}$, 
and $\Xi^{odd\dagger}$, is multiplicatively closed.  
$\Box$ \\ 

In \textbf{[1]} there are seven copies of the octavian integers defined starting from 
$\Xi^{even}$, but needing that strange switching of indices in each case to make the 
set multiplicatively closed ($\Xi^{even}[0-a]$).  We see here that the XY-product can be used to unravel 
the octavian integers so that 
we need only use $\Xi^{even}$, not a rotated copy.  This is reminiscent of the way the 
XY-product unravels triality.  In particular, for all $g \in SO_{8}$ acting on \textbf{O} 
there exist unit elements $X, Y \in \textbf{O}$ (not unique) such that for all $A, B \in \textbf{O}$,
%
%
\begin{equation}
g[A\circ_{XY}B] = g[A]g[B].
\end{equation}
That is, replacing our starting product with the XY-product on the left hand side above means 
we needn't perform triality rotations on $g$ to achieve equality.  
If $g \in G_{2}$, the automorphism group of \textbf{O}, then $X = Y = \pm1$, so $g[AB] = g[A]g[B]$.

As to $\Xi^{even}[0-a]$ being multiplicatively closed, note that if we define $d_{a} = (1 + e_{a})/\sqrt{2}$, then 
$$
\Xi^{even}[0-a] = d_{a} \Xi^{even} d_{a}.
$$
Therefore, exploiting a Moufang identity and the result (12) above, noting that $d_{a} \in \Xi^{odd}$ and 
$d_{a} \in \Xi^{odd\dagger}$, 
$$
\begin{array}{ll}
\Xi^{even}[0-a] \Xi^{even}[0-a]  & = (d_{a} \Xi^{even} d_{a})(d_{a} \Xi^{even} d_{a}) \\
    & = d_{a} ((\Xi^{even} d_{a})(d_{a} \Xi^{even}) )d_{a} \\
    & = d_{a} (\Xi^{even}) d_{a} = \Xi^{even}[0-a].
\end{array}
$$
That is, $\Xi^{even}[0-a]$ are octavian integers as a consequence of a special application of (12).

\section*{5. Integral Quaternions, the Z-Product and $\Lambda_{4}$ Lattices.}

The 4-dimensional quaternion algebra, \textbf{H}, is associative, so for all $A,B,X \in \textbf{H}$, $X \ne 0$,  
$(AX)(X^{-1}B) = AB$.  \textbf{H} hasn't got an X-product like that defined for \textbf{O}.  But the generalization 
of the XY-product leads to the following definition of the quaternion Z-product:
%
%
\begin{equation}
(AX)(Y^{\dagger}B) = AXY^{\dagger}B = A\bullet_{Z}B = AZB, 
\end{equation}
where $Z = XY^{\dagger}$ must be s unit quaternion.  The automorphism group of \textbf{H} is $SO_{3}$.  
A general element of the full $SO_{4}$ group of actions on \textbf{H} 
takes the form
%
%
\begin{equation}
g[A] = UAV^{-1}, 
\end{equation}
with $V^{-1}U$ a unit quaternion.  Let $Z = V^{-1}U$, then
%
%
\begin{equation}
g[A\bullet_{Z}B] = g[A]g[B].
\end{equation}
If $g \in SO_{3}$, then $Z = 1$, and $g[AB] = g[A]g[B]$.  
So the quaternion Z-product bears the same relationship to $SO_{3}$ and $SO_{4}$ as the 
octonion XY-product bears to $G_{2}$ and $SO_{8}$.

Let $q_{m}$, m = 0,1,2,3, be a quaternion basis, with $q_{0}$ the identity.  Define 
%
%
\begin{equation}
\begin{array}{ll}
\Upsilon_{0} = &\{\pm q_{m}\},  \\ 
\Upsilon_{2} = &\{(\pm q_{m}\pm q_{n}\pm q_{r}\pm q_{s})/2: m,n,r,s 
\mbox{ distinct},   \\ 
\Upsilon^{even} =&\Upsilon_{0}\cup \Upsilon_{2}, \\ 
{\cal D}_{4}^{even} =&\mbox{span\{}\Upsilon^{even}\mbox{\}}, \\ \\

\Upsilon_{1} = &\{(\pm q_{m}\pm q_{n})/\sqrt{2}: m,n \mbox{
distinct}\},  \\ 
\Upsilon^{odd} =&\Upsilon_{1}, \\ 
{\cal D}_{4}^{odd} =&\mbox{span\{}\Upsilon^{odd}\mbox{\}} \\ 
\end{array}
\end{equation} \\

Both $\Upsilon^{even}$ and $\Upsilon^{odd}$ have 24 elements, and constitute the inner shells of 
$D_{4}$ lattices.  The elements $\Upsilon^{even}$ are the Hurwitz units of the set of Hurwitz integers (see \textbf{[1]}).  
They are multiplicatively closed.  That being the case, $\Upsilon^{even}$ is multiplicatively closed using 
the Z-product for all $Z \in \Upsilon^{even}$.  If $X,Y \in \Upsilon^{odd}$ and the Z-product with $Z = XY^{\dagger}$ 
preserves quaternion units ($q_{m}\bullet_{Z}q_{n} = \pm q_{k}$), then there exists some index $j$ such that 
$Y = \pm q_{j} X$.  Above we saw that in expanding this unit preserving property from the X-product to the 
XY-product led to an expansion of associated $E_{8} = \Lambda_{8}$ lattices to $\Lambda_{16}$.  Here we get 
an expansion of $D_{4} = \Lambda_{4}$ to $E_{8} = \Lambda_{8}$.  I will not work out the details here.  They 
are similar to the more complicated case developed in \textbf{[6]}.  I conclude this section by noting that $\Upsilon^{even}$ 
is closed under the Z-product for $Z = XY^{\dagger} \in \Upsilon^{even}$.

\section*{6. Octonion Triples and the Leech Lattice.}

Before proceeding we'll define variations on the $\Xi_{k}$:
$$
\begin{array}{ll}
{\mathcal A}_{0} = &\{\pm e_{a}\},  \\ 
{\mathcal A}_{2} = &\{(\pm e_{a}\pm e_{b}\pm e_{c}\pm e_{d})/2: a,b,c,d
\mbox{ distinct},   \\ 
&e_{a}(e_{b}(e_{c}e_{d}))=\pm 1\}, \\ 
{\mathcal A}^{even} =&{\mathcal A}_{0}\cup {\mathcal A}_{2}, \\ \\

{\mathcal A}_{1} = &\{(\pm e_{a}\pm e_{b}): a,b \mbox{
distinct}\},  \\ \\
{\mathcal A}_{3} = &\{(\sum_{a=0}^{7}\pm e_{a})/2:
\mbox{ even number of +'s}  \},   \\ 
&a,b,c,d\in\{0,...,7\},  \\ 
{\mathcal A}^{odd} =&{\mathcal A}_{1}\cup {\mathcal A}_{3}. \\ 
\end{array}
$$
The only change is to the odd elements, which now have rational coefficients $1$ 
and $\frac{1}{2}$.  So these elements are no longer unit octonions, and $\ell_{0} \in {\mathcal A}^{odd}$.  
Still, in general, if $U \in {\mathcal A}_{k}$, any k, then for all basis elements $e_{a}$ and $e_{b}$ there exists 
an $e_{c}$ such that 
%
%
\begin{equation}
(e_{a}U)(U^{-1}e_{b}) = \pm e_{c} = (e_{a}(e_{b}U))U^{-1}.
\end{equation}
The last equality follows from Moufang identities, and it implies by induction that for an arbitrary 
set of octonion units $e_{a}$, $e_{b}$, ... $e_{d}$ there exists a unit $e_{c}$ such that 
%
%
\begin{equation}
e_{a}(e_{b}(...(e_{d}U)...)) = \pm e_{c} U.
\end{equation}
That is, nested products of units from the left on any $U \in {\mathcal A}^{even}\cup{\mathcal A}^{odd}$ collapse 
to a product of a single unit, to within a sign.  In each case, the resulting unit $\pm e_{c}$ depends upon $U$.

In \textbf{[9]} I introduced a representation of the Leech lattice over ${\bf{O}}^{3}$.  Quite frankly, I no longer 
have much of an idea how I arrived at this representation, but more recently Robert Wilson, applying 
more rigorous and reproducible mathematical methods, independently derived a representation of 
$\Lambda_{24}$ over ${\bf{O}}^{3}$ that I will show verifies my initial guess.  

In both our papers the final result breaks up the inner shell of $\Lambda_{24}$, which is of order 196560, into 
three subsets with orders $3\times 240 = 720$, $3\times 240 \times 16 = 11520$, 
and $3\times 240 \times 16 \times 16 = 184320$, the sum of all three orders being 196560.  
The biggest of these subsets is the one I want to focus on.  Translating Wilson's notation into my own, 
the elements of this subset take the form
$$
(\; (P\ell_{0}^{\dagger}) e_{a}, \;\; \pm P e_{c}, \;\; \pm (P e_{a}) e_{c} \; ),
$$
where $P \in {\mathcal A}^{odd\dagger}$, the two $\pm$ signs are independent, the indices $a,c \in \{0,...,7\}$ 
are independent, and to achieve the full subset we include all permutations of these three octonion components.  
As a first step in connecting to my representation we take the conjugate of each of the three components, 
$$
(\; e_{a} (\ell_{0}P), \;\; \pm e_{c} P, \;\; \pm e_{c} (e_{a} P) \; ),
$$
where now it is understood that $P \in{\mathcal A}^{odd}$, and although it may seem like I am playing fast 
and loose with the signs, as long as we have two independent $\pm$ signs on two of these three terms all is ok.
Next we cyclicly permute these three terms to the left, keep the $\pm$ signs on the second two, and 
replace $P$ with $e_{c} P$ (keep in mind, this result takes advantage of sign flexibility):
$$
(\; P, \;\; \pm e_{c} (e_{a} (e_{c}P)), \; \pm  e_{a} (\ell_{0}(e_{c} P)) \; ).
$$
The two $e_{c}$ units appearing in the second component cancel (the resulting sign change 
we absorb into the $\pm$), leaving us with
%
%
\begin{equation}
(\; P, \;\; \pm e_{a} P, \; \pm  e_{a} (\ell_{0}(e_{c} P)) \; ) = 
(\; P, \;\; \pm e_{a} P, \; \pm  (e_{a} \ell_{0} e_{a} )(e_{b} P) \; ), 
\end{equation}
where $e_{b}$ depends on $e_{a}$, $e_{c}$, and $P$.  To within multiplication by a scalar, this is now 
in the form of the representation 
presented in \textbf{[9]} (and I owe a big debt to Robert Wilson for putting that representation on a 
firm mathematical footing).

The remaining elements of this representation of the Leech lattice, also conforming to \textbf{[9]} and 
\textbf{[11]}, take the forms,
%
%
\begin{equation}
(\; 2Q, \;\; \pm 2 e_{a} Q, \; 0 \; ) = (\; \ell_{0}^{\dagger}P, \;\; \pm  e_{a} (\ell_{0}^{\dagger}P), \; 0 \; ), 
\end{equation}
where $Q \in {\mathcal A}^{even}$, or $P =  \ell_{0}Q  \in {\mathcal A}^{odd}$, and we include 
all permutations ($3\times 240 \times 16 = 11520$ elements, and note that 
$(\ell_{0}^{\dagger})^{-1} = \frac{1}{2}\ell_{0}$)), and 
%
%
\begin{equation}
(\; 2P, \;\; 0, \; 0 \; ), 
\end{equation}
where $P \in {\mathcal A}^{odd}$, and we include all permutations ($3\times 240  = 720$ 
elements).

\section*{7. Motivation.}

This paper has ostensibly nothing to do with physics, but my purpose in exploring these ideas does.  
For years I pursued applications of the four division algebras, \textbf{R}, \textbf{C}, \textbf{H} and 
\textbf{O}, to physics, and these efforts met with considerable success.  The dimensions of these 
algebras, 1, 2, 4, 8 (= $2^k, \; k = 0,1,2,3$), are mathematically resonant.  This finite sequence of 
integers is associated with myriad generative mathematical notions.  But this sequence does not 
include the integer 24, nor does it make sense that it should.  There is another finite sequence 
I suggest that is resonant in a different way: 1, 2, 8, 24.  This tidbit from Wikipedia in speaking about 
the Leech lattice:
\begin{quote}"This arrangement of 196560 unit balls centred about another unit ball 
is so efficient that there is no room to move any of the balls; this configuration, 
together with its mirror-image, is the only 24-dimensional arrangement where 
196560 unit balls simultaneously touch another. This property is also true in 
1, 2 and 8 dimensions, with 2, 6 and 240 unit balls, respectively, based on the 
integer lattice, hexagonal tiling and $E_{8}$ lattice, respectively." \end{quote}

As we have seen, $\Lambda_{24}$ can be nicely represented in the 24-dimensional ${\bf{O}}^{3}$.  
Likewise, $\Lambda_{8} = E_{8}$ has a nice representation in ${\bf{H}}^{2}$ (as well as in 
simply \textbf{O}), and $\Lambda_{2}$ and $\Lambda_{1}$ find their most natural expressions in 
 ${\bf{C}}^{1}$ and ${\bf{R}}^{1}$, respectively.  In \textbf{[12]}, inspired by all of this, I took the algebra 
 ${\bf{T}} = {\bf{R}}\otimes{\bf{C}}\otimes{\bf{H}}\otimes{\bf{O}}$, which in part found a roll of my application 
 of the division algebras to physics in my hyper-spinor space ${\bf{T}}^{2}$, and expanded 
 this to 
 $$
 {\bf{T}}^{6} = {\bf{R}}^{1}\otimes{\bf{C}}^{1}\otimes{\bf{H}}^{2}\otimes{\bf{O}}^{3}.
 $$
The spinor space ${\bf{T}}^{2}$ elegantly accounts for one generation (family) of quarks and 
leptons (and their anti-particles), but theoretical consensus puts the total number of families 
at three.  Clearly, if ${\bf{T}}^{2}$ accounts for one family, then ${\bf{T}}^{6}$ would account for 
three, but the dimensionality of ${\bf{T}}^{6}$ is wrong for a conventional spinor space.  I do not 
view this as a deterrence, but as indication that the mathematical tools needed to fully exploit 
${\bf{T}}^{6}$ may not yet be available to us.  My suspicion is that ternary algebras may be 
involved (\textbf{[13]}).  $\Lambda_{8}$, as we have seen, can be given an algebraic structure, 
turning this lattice into the octavian integers.  The question is: can $\Lambda_{24}$ be given 
its own algebraic structure, perhaps involving a ternary multiplication?  No idea yet.

Finally, and to cement the notion that I may be barking mad, I present some possibly coincidental 
numerology thoughts.  The two finite sequences of resonant dimensions given above are 
$$
\begin{array}{rcccc}
n & 1 & 2 & 3 & 4 \\
\frac{1}{2}2^{n} & 1 & 2 & 4 & 8 \\
F_{n} & 1 & 1 & 2 & 3 \\
\frac{1}{2}F_{n}2^{n} & 1 & 2 & 8 & 24 \\
\end{array}
$$
and $F_{n}$ are the first 4 numbers in the Fibonacci sequence, starting from n = 1.

Why should the Fibonacci sequence have anything to do with this 
finite sequence of dimensions associated with special lattices?  No idea, and I wouldn't mention 
it at all except for the following coincidences relating to the sequence of corresponding kissing numbers, 
2, 6, 240, 196560, for the associated lattices $\Lambda_{1}$, $\Lambda_{2}$, $\Lambda_{8}$, $\Lambda_{24}$:
$$
\prod_{k=1}^{3}F_{k} = 2, \; \prod_{k=1}^{4}F_{k} = 6, \; \prod_{k=1}^{6}F_{k} = 240, \; 
\prod_{k=1}^{8}F_{k} = \frac{196560}{3}.
$$
Why 3,4,6,8?  Not sure, although each is 1 plus a prime for the first 4 primes, 2,3,5,7.  So?  Not sure.  
Haven't a clue.  I'd like a clue, but - sadly - I don't have one.

\section*{8. Some thoughts from the original version of this paper.}

The sets $\Xi_{k}$ defined in (4) are associated with our chosen multiplication table defined in (3).  Any change 
in the multiplication table will result in a change in these sets.  Each element $X$ in some $\Xi_{k}$ 
gives rise to an X-product variant of our original multiplication table that takes octonion units to units (see (5)).  
Associated with this new multiplication table will be an altered collection of sets like our original $\Xi_{k}$.  
We will denote these altered sets
$$
^{X}\Xi_{k}.
$$
In particular, the sets $^{X}\Xi_{0}$ and $^{X}\Xi_{1}$ will not be altered, as they are independent of the 
multiplication table.  However, 
$$
^{X}\Xi_{2} = \{(\pm e_{a}\pm e_{b}\pm e_{c}\pm e_{d})/2: a,b,c,d
\mbox{ distinct},   \\ 
e_{a}\circ_{X}(e_{b}\circ_{X}(e_{c}\circ_{X}e_{d}))=\pm 1\}, 
$$
and $^{X}\Xi_{3}$ will be similar to $\Xi_{3}$, with the number of minus signs in the sum being odd or even, 
depending on the product.

One of the remarkable properties of the octonions is that any sum of nested products from the left (or right) 
can be expressed as a sum of nested products from the right (or left).  In particular, for all X in some $\Xi_{k}$, 
and all indices a = 1,...,7,
%
%
\begin{equation}
Xe_{a} = \frac{1}{2}(-e_{a}X + e_{p}(e_{q}X) + e_{r}(e_{s}X) + e_{u}(e_{v}X)),
\end{equation}
where the indices a,p,q,r,s,u,v are distinct, accounting for all the indices from 1 to 7, and 
$e_{p}e_{q} = e_{r}e_{s} = e_{u}e_{v} = e_{a}$.  Using the identity (20) we can rewrite (27):
$$
Xe_{a} = \frac{1}{2}(-e_{a} + e_{p}\circ_{X}e_{q} + e_{r}\circ_{X}e_{s} + e_{u}\circ_{X}e_{v})X.
$$
By (5) this reduces to, 
$$
Xe_{a} = \frac{1}{2}(-e_{a} \pm e_{b} \pm e_{c} \pm e_{d})X = PX.
$$
P must have a norm 1, and the only way for such a linear combination of octonion units to have 
norm 1 is if either $P = \pm e_{m}$ (eg., this certainly occurs if $X = \pm 1$), or 
if all the indices a,b,c,d are distinct.  Because the original set of 7 indices accounted for all 
the indices from 1 to 7, in this latter case the remaining 4 indices above must satisfy
$$
e_{a}\circ_{X}(e_{b}\circ_{X}(e_{c}\circ_{X}e_{d}))=\pm 1.
$$
So in general, 
$$
Xe_{a} = PX, \mbox{ with } P \in \mbox{ } ^{X}\Xi^{even}.
$$

\end{document}